\documentclass[conference,10pt,letterpaper,compsoc]{IEEEtran}
\usepackage{fancyhdr}

\usepackage{microtype} 
\usepackage[capitalise]{cleveref}
\usepackage{graphicx}
\usepackage{enumitem} 
\usepackage{todonotes}
\usepackage{amstext}
\usepackage{fancyvrb}
\usepackage{color}
\usepackage{comment}
\usepackage{url}

\input{sections/pygment}

\graphicspath{{./img/}{./graphs/}}

\newcommand{\ignore}[1]{}
\newcommand{\binom}[2]{{#1 \choose #2}}
\newcommand{\smallbinom}[2]{\footnotesize{#1 \choose #2}}

\title{Rocket: Efficient and Scalable All-Pairs Computations on Heterogeneous Platforms}

\author{\IEEEauthorblockN{
 Stijn Heldens\IEEEauthorrefmark{1}\IEEEauthorrefmark{2},
 Pieter Hijma\IEEEauthorrefmark{2}\IEEEauthorrefmark{3}, 
 Ben van Werkhoven\IEEEauthorrefmark{1},
 Jason Maassen\IEEEauthorrefmark{1},
 Henri Bal\IEEEauthorrefmark{3},
 Rob van Nieuwpoort\IEEEauthorrefmark{1}\IEEEauthorrefmark{2}
}
\IEEEauthorblockA{
 \IEEEauthorrefmark{1}Netherlands eScience Center,
 \IEEEauthorrefmark{2}University of Amsterdam,
 \IEEEauthorrefmark{3}Vrije Universiteit Amsterdam
}
  \{s.heldens, b.vanwerkhoven, j.maassen, r.vannieuwpoort\}@esciencecenter.nl, \{pieter, bal\}@cs.vu.nl
}

\date{\today}

\begin{document}
\maketitle

\begin{abstract}
All-pairs compute problems apply a user-defined function to each combination of two items of a given data set.
Although these problems present an abundance of parallelism, data reuse must be exploited to achieve good performance.
Several researchers considered this problem, either resorting to partial replication with static work distribution or dynamic scheduling with full replication.
In contrast, we present a solution that relies on  hierarchical multi-level software-based caches to maximize data reuse at each level in the distributed memory hierarchy,
combined with a divide-and-conquer approach to exploit data locality,
hierarchical work-stealing to dynamically balance the workload,
and  asynchronous processing to maximize resource utilization. 
We evaluate our solution using three real-world applications (from digital forensics, localization microscopy, and bioinformatics) on different platforms (from a desktop machine to a supercomputer).
Results shows excellent efficiency and scalability when scaling to 96 GPUs, even obtaining super-linear speedups due to a distributed cache.


\end{abstract}


\section{Introduction}
\label{sec:introduction}

All-pairs compute problems, which evaluate a function for each pair of items of a data set,
 are prevalent in many scientific domains including
radio astronomy~\cite{vannieuwpoort2011correlating}, 
microscopy~\cite{heydarian2018templatefree},
bioinformatics~\cite{qi2004whole},
digital forensics~\cite{vanwerkhoven2018jungle},
computer vision~\cite{phillips2005overview}, 
data mining~\cite{maimon2005data},
information retrieval~\cite{zhu2012clustering},
and biometrics~\cite{liu2015iris}.
These problems typically involve calculating some measure, such as the distance or similarity, between pairs of data items, such as images or objects.
%
%
In general, all-pairs compute problems are computationally demanding because of the 
quadratic nature of the problem. 
Additionally, maximizing 
data reuse is necessary to achieve optimal performance, which in turn requires careful consideration of the workload and data distribution.




The coarse-grained parallelism in all-pairs compute problems scales quadratically with the size of the data set.
%
%
%
%
However, many all-pairs applications~\cite{vannieuwpoort2011correlating,heydarian2018templatefree,vanwerkhoven2018jungle}
also exhibit a large amount of fine-grained parallelism, within the pair-wise function, that can be exploited using GPUs.
This makes distributed clusters equipped with GPUs a suitable platform for these applications: pairs can be processed in parallel across the 
different nodes while the computations for each individual pair are parallelized using the GPU.
Since GPUs evolve rapidly,
these clusters
often upgrade in stages throughout their lifetime, leading to highly heterogeneous platforms containing different GPUs.
%

For all-pairs compute problems it is important that we maximize data reuse on all levels in the system since loading an item is expensive as it requires
accessing remote files, unpacking, pre-processing, filtering, and transferring data.
After an item has been loaded, the resulting data should ideally be used for as many pair-wise comparisons as possible.

We see a clear gap in related work when considering workload distribution and data reuse in existing distributed all-pairs compute frameworks.
Some work applies static scheduling assuming the opportunities for data reuse are known in advance~\cite{plimpton1995fast,kleinheksel2018efficient,zhang2015distributed,yeleswarapu2018memorya}, but this approach is not suitable if the pair computations are irregular or if the platform is highly heterogeneous.
Others use dynamic workload distribution combined with full replication to overcome load-imbalance~\cite{moretti2010allpairs,li2013distributed,zhang2016dataaware}, but replicating all data across all nodes is expensive and only feasible for small data sets that fit in local storage.


In this paper, we present a framework called Rocket for efficiently executing all-pairs compute problems on heterogeneous platforms.
Our solution avoids full replication while exploiting dynamic load balancing and works for data sets that do not fit into memory.
We achieve excellent performance by:

\begin{itemize}
\item offering a dedicated system for all-pairs computations that provides a clear separation of concerns between the user's application code and the Rocket runtime system;
\item using a software-based multi-level (distributed) cache to maximize data reuse at all levels;
\item using a divide-and-conquer approach to efficiently exploit data locality, combined with hierarchical random work-stealing to dynamically distribute the workload; and
\item exploiting asynchronous processing to overlap all data movement with useful computation. 
\end{itemize}

We implemented three scientific applications (from digital forensics, localization microscopy, and bioinformatics) using our framework and evaluate their performance on different types of platforms (from a desktop machine to a supercomputer having ${\sim}100$ GPUs).
We propose a performance model to analyze the results and show that Rocket achieves $88.5{-}99.2\%$ efficiency compared to the modeled lower-bound on the run time.
Efficiency increases further when scaling the number of nodes due to a communication scheme that exploits the larger distributed memory capacity, leading to super-linear speedups. 

This paper is structured as follows: Section~\ref{sec:related_work} presents related work, Section~\ref{sec:design} and \ref{sec:implementation} explain the design and implementation of our solution, Section~\ref{sec:applications} describes the three applications, Section~\ref{sec:evaluation} evaluates the performance of our framework, and  Sections~\ref{sec:future} and \ref{sec:conclusion} present future work and conclusions.


\begin{figure}
  \centering
  \includegraphics[width=.45\textwidth]{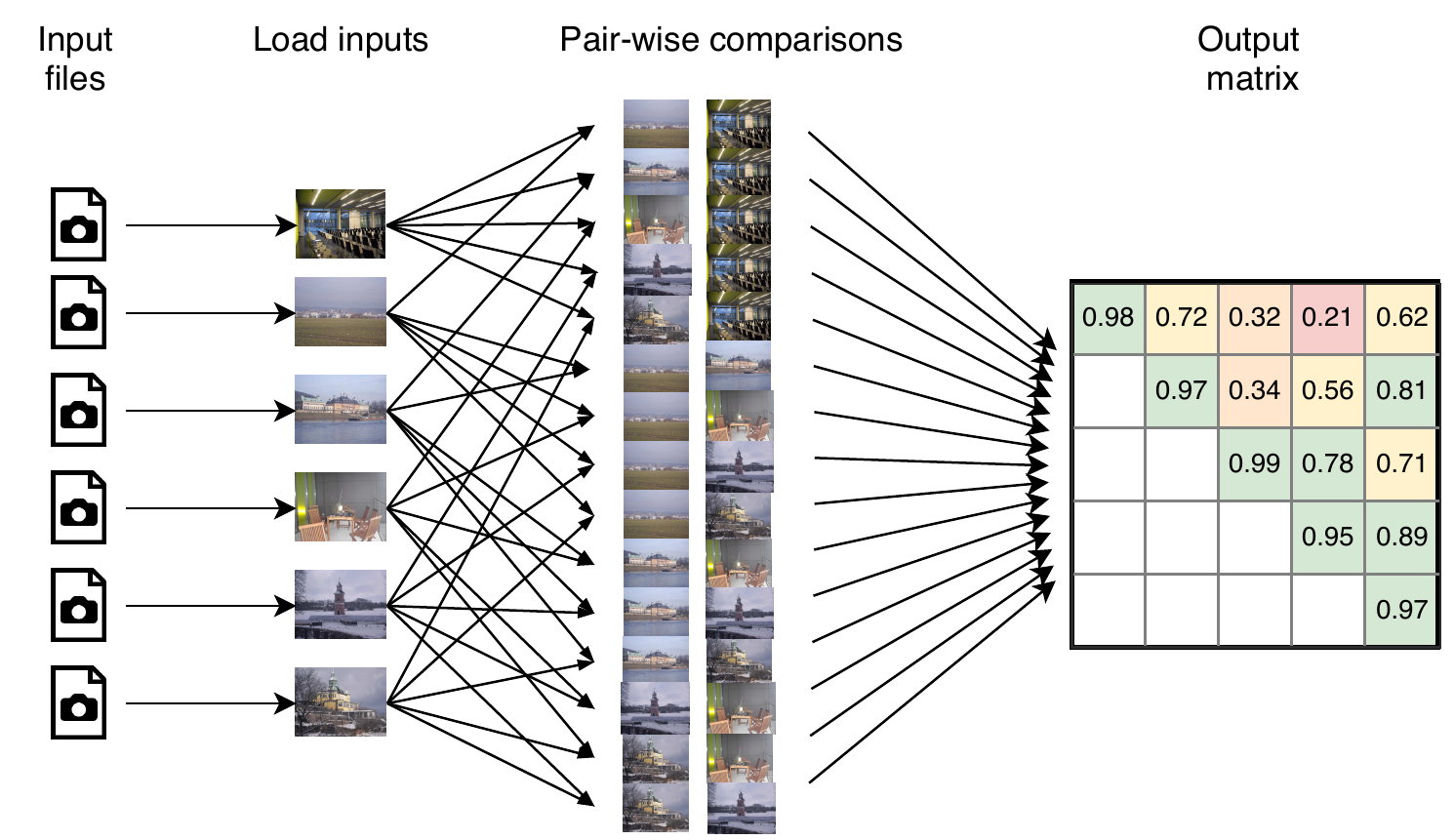}
  \caption{Example of an all-pairs compute problem: calculating the pairwise similarity between images. Photos from the Dresden image database~\cite{gloe2010dresden}.}
  \label{fig:all2all_example}
\end{figure}

\begin{figure*}
  \centering
  \includegraphics[width=.9\textwidth]{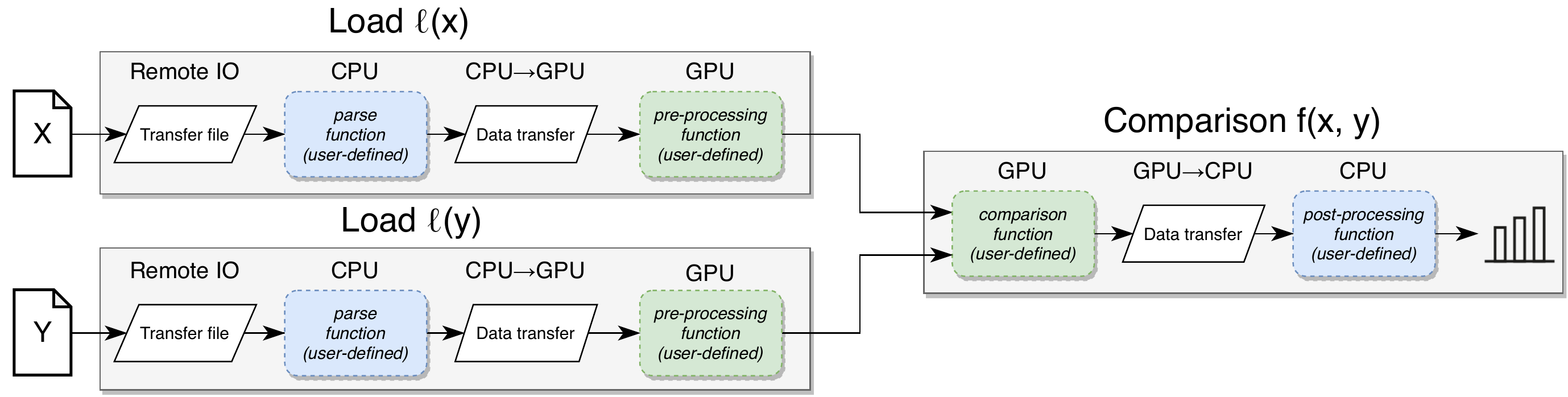}
  \caption{Rocket's pipeline for performing one pair-wise comparison.}
  \label{fig:pipeline}
\end{figure*}



\section{Motivation and Related Work}
\label{sec:related_work}
Several researchers have considered the problem of all-pairs compute problems on distributed systems.
Some have focused on static work distribution where each node is assigned some subset of the pairs to be computed.
This means that the opportunities for data reuse are known in advance.
For instance, processing pairs $(a, b)$ and $(a, c)$ requires loading items $a$, $b$, and $c$ in memory, where item $a$ can be used twice.

Static distribution implies that the $n$ items can be partially replicated since each node can predetermine the subset of items it requires.
The question now becomes how to equally divide the $\binom{n}{2}$ possible pairs over $p$ nodes such that the number of times each item is to be replicated is minimal.
For example, Zhang et al.~\cite{zhang2015distributed} consider all-pairs applications on Hadoop, and they use a heuristic to divide the pairs such that the total computation and data per node is balanced. 
In later work, Zhang et al.~\cite{zhang2016dataaware} reformulate the problem and find a data/work distribution using simulated annealing.
Plimpton~\cite{plimpton1995fast} considered distributed N-body simulations, and they propose a distribution scheme in which each node stores $\frac{2n}{\sqrt{p}}$ items.
Kleinheksel and Somani~\cite{kleinheksel2018efficient} use cyclic quorums to lower this to $\frac{n}{\sqrt{p}}$, which appears to be the best-known lower bound.


Yeleswarapu et al.~\cite{yeleswarapu2018memorya} propose a slightly different static approach having a memory footprint of just $\frac{3n}{p}$.
Each node initially loads $\frac{n}{p}$ inputs in memory and the pair computations are performed in $p$ \emph{rounds}.
Every round involves exchanging items between nodes and processing the pairs that result from combining the local and the received items.

Unfortunately, the above methods all utilize a static workload distribution, which suffers from load-imbalance if the computation is irregular or the platform is heterogeneous. There is also a limit to the size of the problems that can be solved since each node must have sufficient memory to store the assigned items.
Additionally, none of these authors consider systems containing both CPUs and GPUs.


Other researchers have looked into dynamic scheduling of the workload. 
For instance, Zhang et al.~\cite{zhang2016dataaware}  extended their static data distribution scheme to support a limited form of dynamic scheduling.
Their solution is based on the observation that, since data is partially replicated, one pair can sometimes be processed by more than one node in the cluster.
However, while this solution allows some flexibility in scheduling of the work, load imbalance is still a possibility.


Moretti et al.~\cite{moretti2010allpairs} investigate full dynamic scheduling, and they propose \emph{All-pairs}: a framework for all-pairs computation on grids consisting of loosely coupled computers.
Their framework replicates all inputs across all nodes and uses a centralized master to dynamically dispatch batches of jobs.
Li et al.~\cite{li2013distributed} present a similar framework intended for performing pairwise Needleman-Wunsch computations on distributed heterogeneous platforms.
Again, data is replicated across all devices, but they use a two-level scheduling solution: a centralized dispatcher distributes batches of jobs to nodes, and a node-level dispatcher distributes these jobs across the CPUs/GPUs.
However, both solutions require full replication, which is expensive and infeasible if the input data exceeds memory capacity. 

Overall, we observe that all-pairs computation problems show friction between two aspects: workload distribution and data distribution.
Dynamically scheduling the work avoids the problem of workload imbalance, but requires all data to be available at all nodes.
Statically scheduling the work avoids full replication, but could lead to workload imbalance.

In the next sections, we discuss our approach that allows dynamic work distribution while avoiding full replication.
The main differences with related work are that our solution
(1) uses dynamic work scheduling by means of work-stealing without the need for a centralized master, 
(2) supports partial replication by means of caches on multiple levels of the memory hierarchy, and 
(3) fully supports heterogeneous GPU platforms and irregular workloads.


\section{Design}
\label{sec:design}

In this section, we explain the design of our framework. 
The essence of an all-pairs compute problem is straightforward: calculate the result of $f(\ell(i), \ell(j))$ for each pair $(i, j)$ where $1 {\leq} i {<} j {\leq} n$. 
Given are a deterministic function $\ell$ that loads the required data for the $i$-th item into memory and a binary function $f$ applied to the two items. For this work, we assume the inputs are 
coarse-grained static files (e.g., images, sensor data, serialized objects) and $\ell(i)$ reads the $i$-th file and, if needed, parses its content and performs some pre-processing (e.g., filters, transformations, or feature extraction). The results of $x=\ell(i)$ and $y=\ell(j)$ are passed to $f(x, y)$ which computes an application-specific answer, such as a 
correlation score or distance metric. Although $\ell$ and $f$ are presented here as simple functions, they could be processing pipelines that consist of multiple stages 
executed on CPUs and GPUs.

\label{sec:design_pipeline}
For Rocket, we assume the user's processing pipelines for $\ell$ and $f$ follow the pattern shown in \cref{fig:pipeline}.
This design is simple and elegant, making it easy to understand and offering a clear separation of concerns between what the user needs to implement and what is handled by Rocket.
Nevertheless, the model is sufficiently flexible to implement several real-world applications, as we shall demonstrate in Section~\ref{sec:applications}.

The $\ell(i)$ pipeline is performed in four stages:
(1) load the required file from (possibly remote) storage into local memory;
(2) \emph{parse} the file's raw content into the appropriate format on the CPU;
(3) transfer the data from CPU memory to GPU memory;
and, (4) \emph{pre-process} the data on the GPU.

Furthermore, $f(x, y)$ is performed in three stages:
(1) perform the \emph{comparison} on the GPU;
(2) transfer the result from GPU memory to CPU memory;
and, (3) \emph{post-process} the result on the CPU.

\begin{figure}
\footnotesize

\ignore{
interface Application<Key, Result> {
    void getFilePathForKey(Key key);
    void parseFile(Key key,
        HostBuffer fileContents, HostBuffer result);
    void preprocessGPU(Key key,
        DeviceBuffer input, DeviceBuffer output);
    void compareGPU(
        Key leftKey, DeviceBuffer leftItem, 
        Key rightKey, DeviceBuffer rightItem, 
        DeviceBuffer result);
    Result postprocess(HostBuffer result);
}
}

\begin{Verbatim}[commandchars=\\\{\}]
\PY{k+kd}{interface} \PY{n+nc}{Application}\PY{o}{\PYZlt{}}\PY{n}{Key}\PY{o}{,} \PY{n}{Result}\PY{o}{\PYZgt{}} \PY{o}{\PYZob{}}
    \PY{k+kt}{Path} \PY{n+nf}{getFilePathForKey}\PY{o}{(}\PY{n}{Key} \PY{n}{key}\PY{o}{);}
    \PY{k+kt}{void} \PY{n+nf}{parseFile}\PY{o}{(}\PY{n}{Key} \PY{n}{key}\PY{o}{,}
        \PY{n}{HostBuffer} \PY{n}{fileContents}\PY{o}{,} \PY{n}{HostBuffer} \PY{n}{result}\PY{o}{)}\PY{o}{;}
    \PY{k+kt}{void} \PY{n+nf}{preprocessGPU}\PY{o}{(}\PY{n}{Key} \PY{n}{key}\PY{o}{,}
        \PY{n}{DeviceBuffer} \PY{n}{input}\PY{o}{,} \PY{n}{DeviceBuffer} \PY{n}{result}\PY{o}{)}\PY{o}{;}
    \PY{k+kt}{void} \PY{n+nf}{compareGPU}\PY{o}{(}
        \PY{n}{Key} \PY{n}{leftKey}\PY{o}{,} \PY{n}{DeviceBuffer} \PY{n}{leftItem}\PY{o}{,} 
        \PY{n}{Key} \PY{n}{rightKey}\PY{o}{,} \PY{n}{DeviceBuffer} \PY{n}{rightItem}\PY{o}{,} 
        \PY{n}{DeviceBuffer} \PY{n}{result}\PY{o}{)}\PY{o}{;}
    \PY{n}{Result} \PY{n+nf}{postprocess}\PY{o}{(}\PY{n}{HostBuffer} \PY{n}{result}\PY{o}{)}\PY{o}{;}
\PY{o}{\PYZcb{}}
\end{Verbatim}

\caption{The interface that must be implemented by the user for the application.}
\label{vrb:program_interface}
\end{figure}

As an example, consider an algorithm that calculates the pair-wise similarity between photos such as shown in \cref{fig:all2all_example}.
Such an application consists of the following tasks: 
decoding of the image format  on the CPU (\emph{parsing}), 
applying filters to the image on the GPU (\emph{pre-processing}), 
calculating a correlation score on the GPU (\emph{comparison}), 
and applying a threshold to the score on the CPU (\emph{post-processing}).


With this design the user defines four application-specific functions:
parsing (CPU), pre-processing (GPU), comparison (GPU), and post-processing (CPU) adhering to the interface defined in \cref{vrb:program_interface}.  Launching an all-pairs application on the cluster can then be achieved by simply calling Rocket's main class with an input array of \verb=Key= elements.  Rocket will automatically take care of network communication, data transfers, memory management, scheduling, exploiting data reuse, load balancing, and overlapping computation with I/O.

\section{Implementation}
\label{sec:implementation}
This section dives into the implementation of our framework. 
Three aspects lay the foundation for Rocket.
First, a naive approach to processing would be to ignore any form of data reuse and execute both $\ell(i)$ and $\ell(j)$ when processing one pair $(i, j)$.
However, this is expensive since the cost of loading the items, often considerably outweighs the cost of actually comparing the two items.
Fortunately, the functions $\ell(i)$ and $\ell(j)$ are deterministic meaning that, after they have been executed once, their results can reused for future jobs that require item $i$ or $j$. 
Exploiting this data reuse will significantly improve performance. 
We use a software-based cache to store these results at different levels of the distributed memory hierarchy. 
This design is described in \cref{sec:design_cache}.

Second, as discussed in \cref{sec:introduction}, dynamically scheduling the workload is necessary to avoid the problem of load imbalance, either due to irregular completion time of $\ell$ or $f$, or because of heterogeneity of the platform. 
A straightforward solution would be to have a single master dynamically distributing the pairs $(i, j)$ to worker nodes.
However, this would not take data locality into account which is crucial to maximize cache effectiveness.
Moreover, scalability would be limited due to having a central point. 
Instead, we have chosen to perform load-balancing by a divide-and-conquer approach with hierarchical work-stealing, since it shows excellent scalability and data locality in practice. 
This allows for dynamic load balancing while also exploiting data reuse.
This solution is described further in \cref{sec:design_scheduling}.

Third, to maximize resource utilization, Rocket keeps a large number of comparison jobs in progress at all times and relies on asynchronous processing to make progress.
This approach maximizes throughput (i.e., number of comparisons performed per second) and thus minimizes the runtime of the all-pairs application.
\Cref{sec:design_asynchronous} describes this design.

We have implemented Rocket using 
Ibis~\cite{vannieuwpoort2002ibis} as communication library,
Constellation~\cite{maassen2011junglea} for distributed work-stealing,
CUDA for GPU programming, and 
the Xenon library~\cite{maassen2020xenon} to access remote storage resources.

\subsection{Multi-Level Caching}
\label{sec:design_cache}

\begin{figure}
  \centering
  \includegraphics[width=\columnwidth]{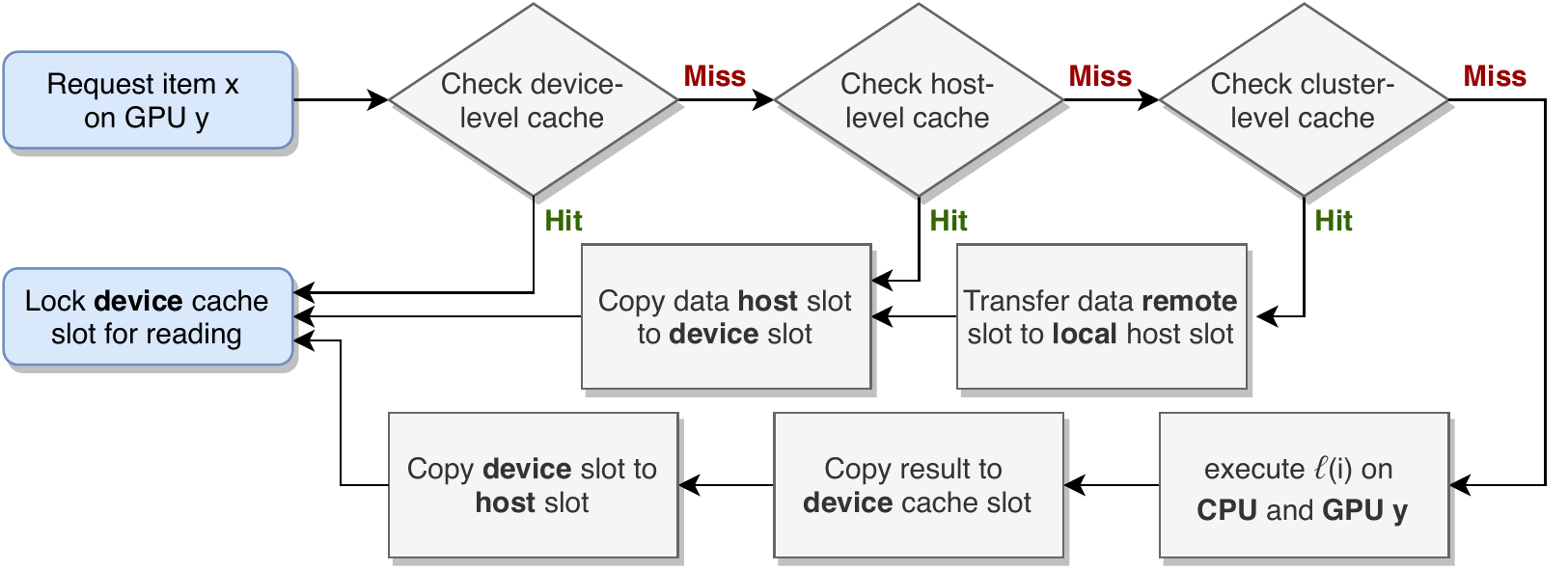}
  \caption{Flow diagram describing the cache policy.}
  \label{fig:cache_diagram}
\end{figure}

As explained before, exploiting data reuse is essential since re-executing the entire pipelines of $\ell(i)$ and $\ell(j)$  for each pair $(i, j)$ is expensive.
For instance, our evaluation presents an application from digital forensics (\cref{sec:evaluation}) where the average GPU compute time is $1$ ms for one comparison, but parsing one input file takes $130$ ms.
By storing the results of each execution of $\ell$ in a cache, the number of times items need to be loaded is reduced and the system can be fully dedicated to performing comparisons.

Rocket uses a three-level software-based cache to maximize the available memory capacity by storing results at different levels of the memory hierarchy.
At the first level is a per-device cache that stores the results in GPU memory since both the last stage of $\ell$ and first stage of $f$ are performed on the GPU.
At the second level is a per-node host memory cache that extends the first-level device cache with the usually much larger memory capacity (${\sim}10\text{-}100$~GB) of the host compared to that of an average GPU (${\sim}5\text{-}10$~GB).
At the third level is a cluster-wide communication scheme that allows nodes to query remote caches, essentially establishing one large distributed memory cache.
Note that these caches are managed in \emph{software} and should not be mistaken with \emph{hardware} caches such as disk caches, L1/L2/L3 cache on CPUs, or shared memory on GPUs.


Below we describe each level in detail. 
\Cref{fig:cache_diagram} visualizes how these different levels interact.

\subsubsection{First-level (Device)} 
At the first level is a per-device cache that manages a fixed number of fixed-sized slots. This cache resides in GPU device memory.
Each slot contains a memory buffer and a status flag which can be \verb=WRITE= (a writer is active) or \verb=READ= ($n$ readers are active).

When a job $(i, j)$ is submitted, this cache is checked for items $i$ and $j$.
On cache miss for item $i$ (or $j$),  the least-recently-used slot is evicted (discarding its content) and assigned to item $i$ after which the slot is set to \verb=WRITE=.
Now the result of $\ell(i)$ needs to be copied into the slot from the next level cache after which the flag is set to \verb=READ=.

On cache hit, the status flag is checked.
For \verb=READ=, the comparison $f$ can be performed immediately (assuming $j$ is also available) while the number of readers is temporarily incremented.
For \verb=WRITE=, another job is busy writing to the slot so the current job is put on hold until the data becomes available.
Note that the cache thus introduces synchronization between jobs: while one job is writing item $i$, other jobs that depend on item $i$ are stalled until the slot becomes available.
In practice, this does not lead to performance issues, because Rocket ensures that a sufficient number of concurrent jobs are in progress at all times (see Section~\ref{sec:design_asynchronous}).

\subsubsection{Second-level (Host)} 
At the second level is a per-node cache that stores the results in page-locked main memory.
The implementation is similar to that of the device cache, only buffers reside in main memory instead of device memory.
This cache is thus shared by all GPUs in one node.

On a first-level device cache miss, the second-level host cache is checked for the item.
On a hit, the contents of the host slot is transferred to the device slot. 
As we show later (see \cref{sec:design_asynchronous}), the overhead of copying data between host and device caches is negligible since Rocket overlaps data transfers and computation.
On a miss, an item is evicted and the empty slot is assigned to item $i$ for which the data needs to be obtained from the next level cache.
Note that data is thus always written to both the device and host cache. 
We chose this solution since it is important for the third-level cache that allows nodes to query remote host caches.

\subsubsection{Third-level (Distributed)}
\label{sec:design_cache_distributed}
At the third level we use a communication scheme that, after a local cache miss, allows nodes to query the host cache of remote peers.
While the previous two levels reduce the number of loads per \emph{node}, the third level reduces the loads for the \emph{cluster as a whole}.

An important consideration is how to locate a node that has the required data in its local cache, which is non-trivial since we use dynamic scheduling.
A centralized registry that keeps track of the data on nodes would be a poor solution since it introduces a central bottleneck and requires excessive coordination and bookkeeping.
Alternatives we considered were broadcasting the request to all peers (not a scalable solution) or to one (or several) randomly chosen peers (but the probability that a random node has one specific item is slim).


Instead, we use a simple communication scheme that allows the system to form a distributed hash table. 
Our scheme is based on the observation that a node that requested an item in the past, will eventually find the data and keep it for some time into the future.
Each node is assigned the responsibility to serve as \emph{point-of-contact} for some subset of the items, where item $i$ is assigned to node $i~\textit{mod}~p$ and $p$ is the number of nodes.
These nodes do not necessarily store these items themselves.
Instead, each node keeps track of a local bookkeeping array \verb=candidates= where \verb=candidates[i]= stores the list of the $h$ nodes that most recently requested item $i$ in the past and are considered the most likely \emph{candidates} for future requests.

For node $A$ to request an item, the following steps are taken:

\begin{itemize}
\item Node $A$ sends a request for item $i$ to node $B=i~\textit{mod}~p$.

\item Node $B$ retrieves from \verb|candidates[i]| the nodes $C_1,...,C_h$  and prepends $A$ to \verb|candidates[i]|.
The request and the list $C_2, ..., C_h$ are forwarded to node $C_1$.

\item Each node $C_x$ checks its local host cache for item $i$:

\begin{itemize}

\item On a hit, it sends the data directly to node $A$.

\item Otherwise, if $x < h$, it forwards the request to $C_{x+1}$.

\item Otherwise, it sends a failure directly to node $A$.

\end{itemize}
\end{itemize}

If this best-effort mechanism does not provide the item, node $A$ is forced to execute $\ell(i)$ locally.
In essence, node B acts as a mediator helping node A (searching the data) to find node C (offering the data).
The parameter $h$ determines the maximal number of \emph{hops} to check, and we evaluate this parameter in \cref{sec:evaluation}.
This scheme is scalable, contains no central component, requires a small amount of bookkeeping (only the \verb=candidates= array) and communication (just $h+2$ messages per request).
Note that in some scenarios, node B or node $C_{x}$ could be node A itself, but this does not affect the correctness of the scheme.


\begin{figure}[t!]
  \centering
  \includegraphics[width=0.4\textwidth]{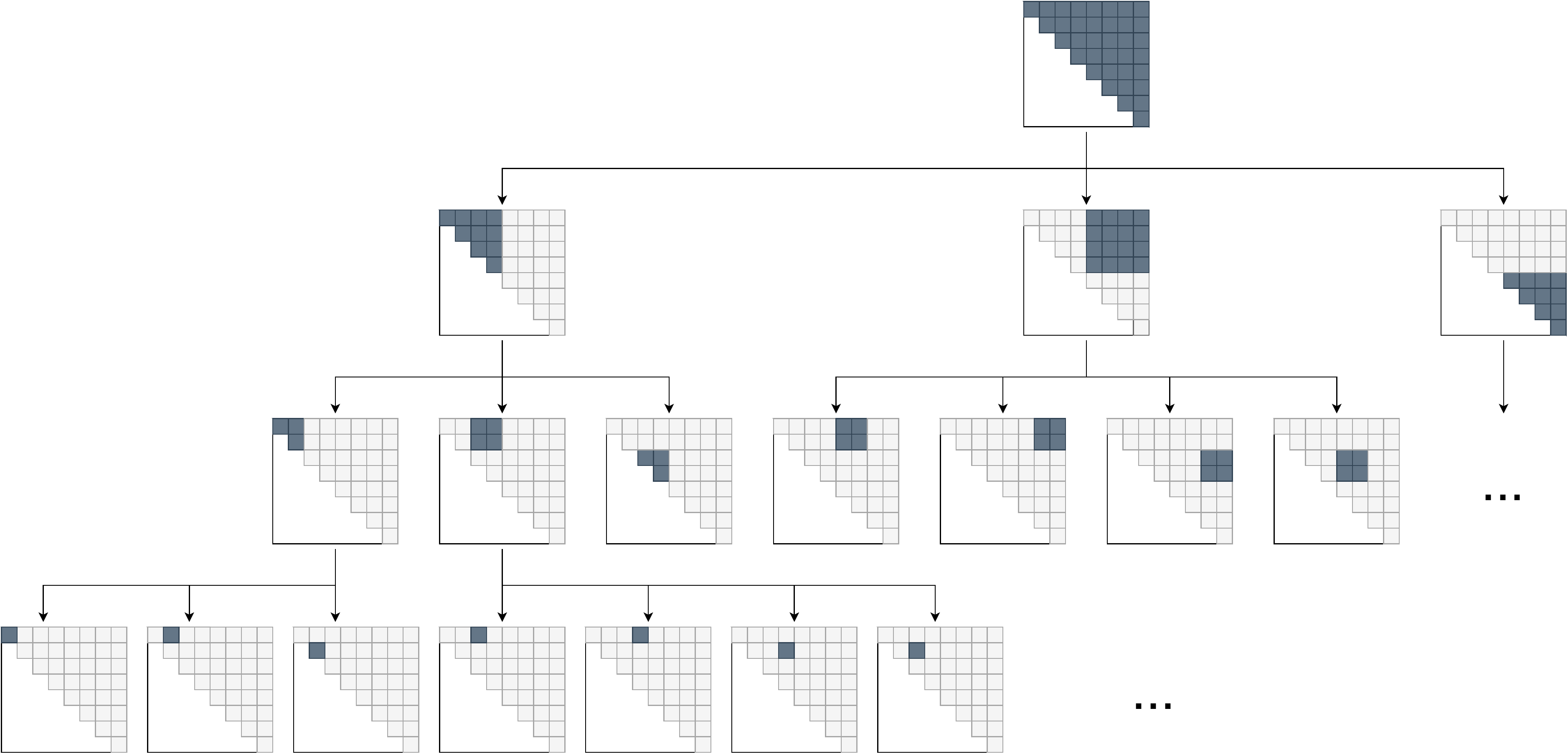}
  \caption{Hierarchical splitting an all-pairs workload of $8{\times}8$ items.}
  \label{fig:hierarchical_scheduling}
\end{figure}

\subsection{Locality-Aware Work Scheduling}
\label{sec:design_scheduling}

To dynamically schedule the pairs $(i, j)$, Rocket uses a divide-and-conquer approach together with work-stealing inspired by frameworks such as Cilk~\cite{blumofe1996cilka} and Satin~\cite{nieuwpoort2010satin}. 
Divide-and-conquer is a common technique in which a larger problem is recursively divided into smaller sub-problems until they become small enough to compute directly. 
It is known that this approach naturally offers excellent data locality while allowing for dynamic workload balancing~\cite{blelloch2008}.

Recall that the total workload consists of processing each pair $(i, j)$ where  $1 {\leq} i {<} j \leq n$.
This workload can be seen as an upper triangular matrix.
This larger matrix can be split into four sub-matrices, one for each quadrant, and each sub-matrix can recursively be split into smaller quadrants until eventually reaching individual entries $(i, j)$.
\Cref{fig:hierarchical_scheduling} shows this process for a small $8\times8$ matrix.
Note that quadrants may sometimes contain no work; these can be ignored.

Rocket performs distributed work-stealing using \emph{Constellation}~\cite{maassen2011junglea}: a software platform for distributed, heterogeneous, hierarchical environments.
During initialization, each node launches one Constellation worker thread per GPU.
The master node then spawns a single root task representing the entire matrix to be computed.
This task spawns four new tasks (one for each quadrant of the matrix) and each sub-task recursively spawns four new tasks, one for each sub-quadrant.
The tasks at the lowest level represent a single $(i, j)$ entry and these leaf tasks submit the actual job $(i, j)$ to the Rocket runtime system.
Worker threads always prioritize local tasks at the lowest level in the tree since these provide the best data locality.

Load balancing is performed by \emph{random work-stealing}: Workers that become idle will repeatedly attempt to steal a task from a randomly selected peer. 
This technique has been shown to be a suitable solution to balance workload in distributed environments~\cite{nieuwpoort2010satin}.
The task stolen is always at the `highest' level (i.e., the largest task available) since it results in the most work per steal request. 
Stealing is performed hierarchically: workers first attempt to steal from a worker on the same node before selecting a remote node. 
The advantage of work-stealing over master-worker is that it exhibits good data locality while also balancing the workload: if there are no idle nodes, work is not stolen and thus executed locally on the node that generated it. 
It is well-known that divide-and-conquer leads to excellent exploitation of locality, both for hierarchical memory architectures~\cite{blumofe1996cilka} and in distributed systems~\cite{nieuwpoort2010satin}.

Rocket's runtime system operates asynchronously and submitting a job does not block the caller.
Without any form of back-pressure, one node could rapidly claim all available work meaning others become idle.
To prevent this, Rocket has a \emph{concurrent job limit} parameter that limits the number of concurrent jobs that can be simultaneously submitted to Rocket. 
Once this limit is reached, worker threads will stop submitting new jobs until an older job completes. 

\subsection{Asynchronous Processing}
\label{sec:design_asynchronous}

\begin{figure*}[t!]
  \centering
  \frame{\includegraphics[width=\textwidth]{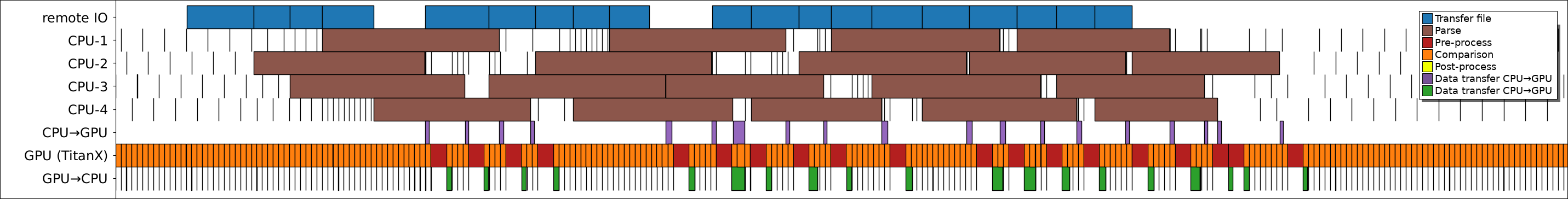}}
  \caption{Small section of a trace from the forensics application (\cref{sec:applications}) visualized on a timeline. Rows represents threads and boxes represent executed tasks.}
  \label{fig:gantt}
\end{figure*}

A naive implementation to process submitted pairs is to have one thread (or a small pool of threads) processing the submitted pairs synchronously one-by-one.
However, this would lead to inefficient resource usage since it can result in moments of resource contention (e.g., all threads could perform I/O simultaneously) while under-utilizing other resources (e.g., the GPU is idle in the meantime). 

To maximize resource utilization, Rocket keeps many jobs in progress (\emph{concurrent job limit} as described in \cref{sec:design_scheduling}) and relies heavily on asynchronous processing to make progress on these jobs.
Different threads are launched during initialization with each thread responsible for one type of resource, meaning that tasks executed by different threads do not interfere with each other.
For our current implementation, Rocket launches the following types of threads:

\begin{itemize}[leftmargin=60pt,itemsep=0pt]
\item[\textbf{CPU}] A thread pool performs CPU computations.
\item[\textbf{GPU}] One thread per GPU to launch device kernels.
\item[\textbf{CPU$\rightarrow$GPU}] One thread per GPU for performing data transfers from host to device memory.
\item[\textbf{GPU$\rightarrow$CPU}] One thread per GPU for performing data transfers from device to host memory.
\item[\textbf{I/O}]  One thread for I/O on the (remote) file system.
\end{itemize}

With this scheme, CPU processing, GPU processing, data transfers, and I/O operations are all overlapped.
For example, multiple \emph{parsing} tasks can be executed simultaneously on CPUs together with a \emph{comparison} task running on the GPU, while also transferring data to and from GPUs and performing I/O operations in parallel.
An optional profiling flag can be enabled to trace the tasks executed by different threads, which can be useful for debugging purposes and performance analysis.
\Cref{fig:gantt} shows an example of such a trace visualized on a timeline.

Asynchronous processing is essential for Rocket to achieve good performance since it means that resources are fully utilized.
For instance, for each job $(i, j)$, a first-level cache hit on items $i$ and $j$ is cheap (since the comparison pipeline can be executed immediately), but a cache miss is expensive since it involves many steps (e.g., data transfers, I/O, parsing, pre-processing, etc.).
It is thus important to keep a larger number of jobs active so that the system can `anticipate' first-level cache misses and acquire the necessary data before running out of work.
\Cref{fig:gantt} demonstrates this well: the GPU remains fully utilized since sufficient work is available, even though slow I/O and CPU tasks are performed in the background.



\section{Applications}
\label{sec:applications}
To demonstrate the generality of our framework, we use three scientific applications that are used by researchers in digital forensics, localization microscopy and evolutionary biology. 
These are realistic applications, not simplified benchmarks, and include all required pre-processing, I/O, and application logic.
The computational kernels are taken as black boxes and are not analyzed in this work.
The applications have different compute and data characteristics, thus demonstrating the generality of our approach.


\ignore{
\begin{table*}
\centering
\begin{tabular}{l||l|l|l}
& C & M & G \\
\hline\hline
Parse & Decode JPEG & Parse JSON & Decompress FASTA file \\\hline
Preprocessing & Extract PRNU pattern & N/A & Build Composition Vector \\\hline
Comparison & Compute PCE & Iterative optimization & Compute cosine similarity \\\hline
Post-processing & N/A & N/A  & N/A \\
\end{tabular}
\caption{}
\end{table*}
}

\subsection{Common-Source Identification (Forensics)}

Common-source identification is a digital forensics application that takes a set of images and identifies which images were made by the same camera
based on sensor noise patterns.
These noise patterns, called \emph{Photo Response Non-Uniformity} (PRNU) patterns~\cite{fridrich2013sensor}, originate from small
deficiencies in the imaging sensor, resulting in small differences in the responsivity. In the resulting images, this leads to
pixels being brighter or darker while having received equal saturation.
To find the images that have been acquired by the same sensor, the noise patterns of all images have to be extracted and compared with each other.

Our Rocket-based implementation is based on the application by Van Werkhoven et al.~\cite{vanwerkhoven2018jungle} developed for the Netherlands 
Forensics Institute. 
We reuse their GPU kernels for extracting the PRNU patterns from images and for computing the similarity between the PRNU patterns of different images.
The metric that is used to measure the similarity of two PRNU patterns is the \emph{Normalized Cross Correlation}.
%
The decoding of the JPEG format is done on the CPU using \verb=libjpeg=.
The application compares images that have equal dimensions and as such, computations are highly regular.

Our data set consists of images having dimensions $3648\times{}2736$ from the Dresden image database~\cite{gloe2010dresden}, which is developed
specifically for the goal of researching PRNU-based algorithms.


\subsection{Phylogeny Tree Construction (Bioinformatics)}

Phylogenetic tree reconstruction is the problem of reconstructing how species descend from common ancestors given their genetic material.
The popular alignment-free method by Qi et al.~\cite{qi2004whole} is a fast algorithm to achieve this by hierarchical clustering of the distance matrix between all species.
The distance between two species is calculated based on the distance of their \emph{composition vectors} (CVs).
The CV of a species is derived from the frequency of substrings, of a chosen length $k$, in their protein sequences.
These CVs are represented as sparse vectors and have between 100.000 and 1.800.000 entries.
Extracting these CVs is expensive since it requires scanning the entire genome, but comparing two CVs is cheap, essentially being the dot product between two sparse vectors.
Computation is irregular since the vectors are sparse.

We have implemented this algorithm in CUDA based on the description and formulas by Qi et al.~\cite{qi2004whole}.
The input data set consists of 2500 randomly chosen reference bacteria proteomes from Uniprot Proteomes database~\cite{2019uniprot} and files are stored in compressed FASTA format.
Pre-processing consist of decompressing the file (on CPU) and constructing the CV (on GPU), while comparing two CVs is done entirely on the GPU.


\subsection{Localization Microscopy Particle Fusion (Microscopy)}

Localization microscopy is an optical super-resolution microscopy method that operates on the localization of individual fluorophores, rather than
pixelated images, obtained by a fluorescence microscope. To achieve a resolution well beyond the diffraction limit, multiple images of the same structure
are fused to improve the signal-to-noise ratio and the resolution. 
The method by Heydarian et al.~\cite{heydarian2018templatefree} uses {\em all-to-all registration} of particles to achieve robustness against individual 
misregistrations and under labeling.

Our Rocket-based implementation reuses the GPU kernels from the application by Heydarian et al.~\cite{heydarian2018templatefree}.
These kernels implement two different methods to score the similarity of two particles, which in turn consist of thousands of localizations.
%
The first method is a quadratic L2-distance metric between two Gaussian Mixture Models~\cite{jian2010robust} and the second method is
known as the Bhattacharya distance function~\cite{heydarian2018templatefree}.
%
An optimizer calls these two methods many times and therefore
the
registration process is very compute-intensive, even on a small number of localizations, and heavily data-dependent, making the execution time highly irregular.

Our benchmark data set was generated using the simulator by Heydarian et al. and contains 256 particles stored in JSON format.
Each particle consists of between 1000 and 2000 localizations. 
Since the application works directly on the localizations, there is no pre-processing required other than reading and parsing the particle files.



\section{Experimental Evaluation}
\label{sec:evaluation}
In this section, we evaluate our framework's performance.
After discussing a basic performance model and our experimental setup, we analyze results for one node, a homogeneous cluster, a heterogeneous cluster, and Cartesius (the Dutch National supercomputer).

\subsection{Performance Model}
To establish a baseline for the performance of Rocket, we present a performance model which determines a lower bound on the run time using a hypothetical computing system.

Given $n$ items, the \emph{comparison} pipeline (\cref{fig:pipeline}) must be executed $\smallbinom{n}{2}=\frac{n^2-n}{2}$ times (once for each pair).
The \emph{load} pipeline must be executed at least $n$ times (once for each item), but it may be executed more than $n$ times (i.e., items were evicted from cache). 
We assume $R\,n$ loads are performed in total, where $R$ indicates the number of loads relative to the data set size.
For instance, $R={4.3}$ indicates that each item was, on average, loaded 4.3 times.

$R$ serves  as a basic metric for data reuse since a lower value indicates fewer loads and thus better reuse of previously loaded items.
With perfect data reuse, each item is loaded once (thus $R=1$). In practice $R>1$ due to two reasons: (1) insufficient local cache capacity means that items are evicted that are later loaded again and (2) different nodes in a distributed environment load the same item independently of each other.

For simplicity, we assume our system contains one CPU and one GPU. 
The total GPU processing time ($T_\text{GPU}$) is determined by executing pre-processing $R\,n$ times and comparison $\smallbinom{n}{2}$ times  (where $t_x$ is the average execution time of stage $x$):

\begin{equation}
T_\text{GPU} = R\,n\ t_\text{pre-process} + \binom{n}{2}\,t_\text{comparison}
\end{equation}

The total CPU processing time ($T_\text{CPU}$) is determined by executing parsing $R\,n$ times and post-processing $\smallbinom{n}{2}$ times. 

\begin{equation}
T_\text{CPU} = R\,n\ t_\text{parse} + \binom{n}{2}\,t_\text{post-process}
\end{equation}

The time spent on I/O can be estimated based on the file sizes and the average I/O bandwidth.
However, the actual bandwidth depends heavily on the load on the storage system.

\begin{equation}
T_\text{IO} \approx R\,n\ \frac{\small\text{average~file~size~in~MB}}{\small\text{IO~bandwidth~in~MB/sec}}
\end{equation}

Overhead of CPU-GPU data transfers is negligible since it is easily overlapped.
Perfectly overlapping CPU time, GPU time, and I/O means the total run time will be the maximum of $T_\text{CPU}$, $T_\text{GPU}$, and $T_\text{IO}$.
This motivates why it is important to maximize data reuse: $R$ appears in all three equations and thus minimizing $R$ means maximizing performance.

To establish a baseline, we assume our system has infinite memory and thus perfect data reuse (i.e., $R=1$), I/O has infinite bandwidth (i.e., $T_{IO} \approx 0$), and most processing is performed on the GPU ($T_\text{GPU} > T_\text{CPU}$).
In this scenario, the lower bound on the runtime $T_\text{min}$ equals $T_\text{GPU}$ for $R=1$.

\begin{equation}
T_\text{min} = n\ t_\text{pre-process} + \binom{n}{2}\,t_\text{comparison}
\end{equation}

Our system would show optimal performance on $p$ nodes if the measured run time is $T_\text{min}/p$.
We therefore define \emph{system efficiency} as the ratio between this modeled lower bound on the runtime and the actual measured run time $T$ on $p$ nodes.

\begin{equation}
\text{system efficiency} = \frac{T_\text{min} / p}{T}.
\end{equation}

\begin{table}[t]
\caption{Characteristics of applications for NVIDIA TitanX Maxwell. Time is reported as average $\pm$ standard deviation.}
\label{tab:experiment_baseline}
\centering
\resizebox{\columnwidth}{!}{
\begin{tabular}{l||r|r|r}
Application & Forensics & Bioinformatics & Microscopy \\ 
\hline
\hline
No. of input files ($n$) & 4980 & 2500 & 256 \\
{Size of raw data on disk} & {19.4~GB} & {1.8~GB}  & {150~MB} \\
{Size of preprocessed data in memory} & {189.7~GB} & {110.0~GB} & {0.7~MB} \\
\hline
No. of pairs & 12,397,710 & 3,123,750 & 130,816 \\
{Total data pair-wise processed} & {944.7~TB} & {275.0~TB} & {179.2~MB} \\
\hline
Cache Slot Size & 38.1~MB & 145.8~MB & 6.0~KB \\
No. Device Cache Slots & 291 & 81 & 256 \\
No. Host Cache Slots & 1050 & 280 & 256 \\
\hline
Time Parse (CPU) & 130.8$\pm$14.11 ms & 36.9$\pm$14.79 ms & 27.4$\pm$1.56 ms \\
Time Pre-process (GPU) & 20.5$\pm$0.02 ms & 27.0$\pm$4.90 ms & N/A \\
Time Comparison (GPU) & 1.1$\pm$0.01 ms & 2.1$\pm$0.79 ms & 564.3$\pm$348 ms \\
Time Post-process (CPU) & 0 ms & 0 ms & 0 ms \\
\hline
\end{tabular}}
\end{table}

\begin{figure}[t]
  \centering
  \includegraphics[width=.95\columnwidth]{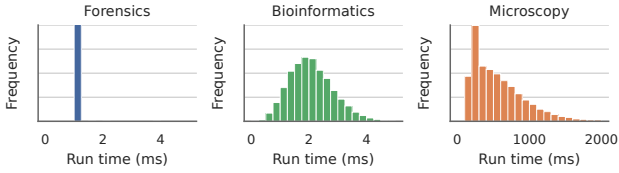}
  \caption{Histogram of the run times for the comparison kernel (i.e., $t_\text{comparison}$) from the three applications. Note the different scales on the horizontal axis.} 
  \label{fig:experiment_times}
\end{figure}

\subsection{Experimental Setup}
Experiments were performed on DAS-5~\cite{bal2016mediumscale}, the distributed platform for experimental computer science research in the Netherlands.
We used the VU site; each node has two Intel Xeon E5-2630 CPUs (16 cores total), offers 64~GB of memory (40~GB allocated to host cache), runs CentOS Linux~7, and nodes are connected by 56~Gb/s InfiniBand FDR.
The site offers a variety of GPUs across the different nodes.
We use MinIO over InfiniBand to serve as a central file storage server.

The last section discusses experiments performed on the Dutch national supercomputer (\emph{Cartesius}\footnote{\url{https://www.surf.nl/en/dutch-national-supercomputer-cartesius}}).
Each node has two E5-2450 v2 CPUs (16 cores total), two NVIDIA Tesla K40m GPUs, offers 96~GB of main memory (80~GB allocated to host cache), and two Mellanox ConnectX-3 InfiniBand adapters (each providing 56~Gb/s inter-node bandwidth).

\subsection{Single Node}
\label{sec:evaluation_single}

\Cref{tab:experiment_baseline} shows information on the data set size for the three applications including the size on disk (i.e., the total of the compressed input files) and the size in memory (i.e., the total after parsing and pre-processing each file).
For the forensics application and the bioinformatics application, the data set increases considerably in size after pre-processing and does not fit into memory of a single node.
For the microscopy application, the data set size is small and actually decreases during pre-processing due to the conversion from a text-based to a binary format.
The table also shows the total amount of data that needs to be combined to perform all pair comparisons (i.e., each of the $n$ items is retrieved $n$ times), highlighting the quadratic nature of all-pairs compute problems.
For instance, for the forensics application, this total reaches almost 1 PB.

To establish a baseline, we ran the three applications on one node equipped with one NVIDIA TitanX Maxwell.
\Cref{tab:experiment_baseline} shows the timing results and cache configuration for these runs.
The table clearly shows that the three applications have different compute- and data-characteristics:
The microscopy application is compute-intensive (comparisons are slow and a small amount of data is processed) while the other two applications are data-intensive (comparisons are fast and a large amount of data is processed).
The forensics application is slightly more data-intensive than the bioinformatics application since more data is processed and comparisons are faster.
For all three applications, the post-processing stage on the CPU is negligible since it only interprets the GPU's result.

\Cref{fig:experiment_times} shows the distribution of run times of one comparison and confirms that the forensics application is a regular problem, while the other two are highly irregular.
Dynamic load balancing for those applications is thus necessary since static scheduling could lead to load-imbalance.

\begin{figure}[t]
  \centering
  \includegraphics[width=.95\columnwidth]{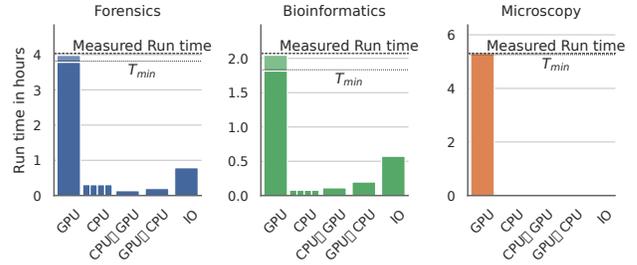}
  \caption{Processing time per thread for each application using one node (TitanX Maxwell). The GPU bar is divided into \emph{pre-processing} (top) and \emph{comparison} (bottom). The dashed line indicates the start-to-end run time of the framework while the dotted line indicates the value of $T_\text{min}$.}
  \label{fig:experiment_baseline}
\end{figure}

\Cref{fig:experiment_baseline} shows, for each application, the overall run time of Rocket together with the total active time of each thread.
The data per thread was extracted from a profile trace by taking the total time of tasks executed by each thread.
The figure shows that all three applications are GPU-intensive since the GPU processing time is dominant. 
Additionally, the results show that the overall run time of the framework equals the GPU processing time, indicating that the asynchronous processing excellently overlaps GPU processing with other activities in the system.
For instance, the bioinformatics application spent more than 30 minutes on I/O operations, but this had no impact on the overall run time since it is overlapped with GPU processing.

The system efficiencies are high: 94.6\% (forensics), 88.5\% (bioinformatics), and 99.2\% (microscopy).
They would increase even further if more memory were available which would allow better data reuse and would lower the overhead of loading items multiple times.
\Cref{tab:experiment_baseline} indicates that, for two applications, only a fraction of the inputs can be cached in host memory slots (21.1\% for forensics and 11.2\% for bioinformatics).

\begin{figure}[t]
  \centering
  \includegraphics[width=.95\columnwidth]{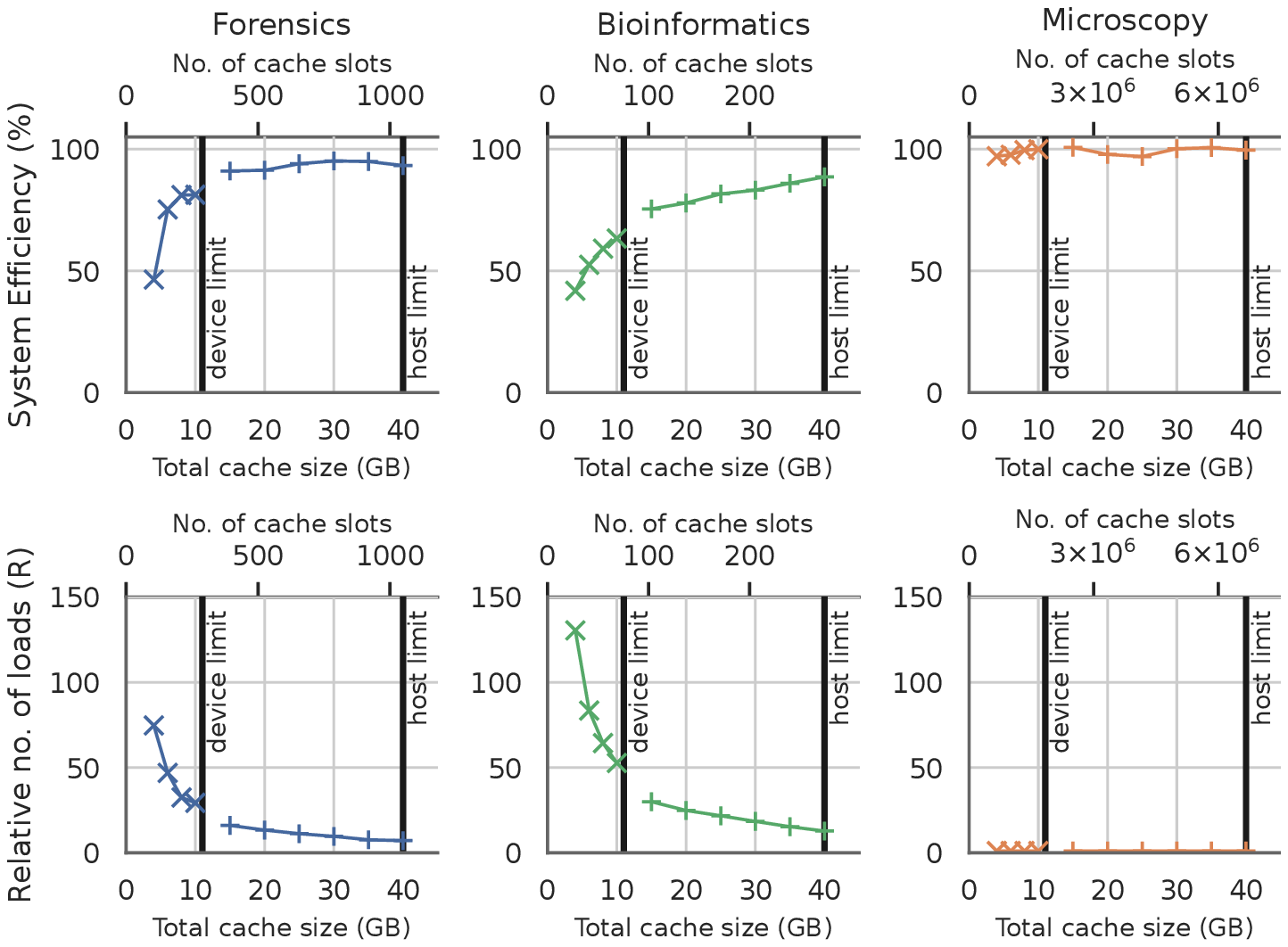}
  \caption{System efficiency and factor $R$ versus cache size on one node (NVIDIA TitanX Maxwell).}
  \label{fig:experiment_cache_size}
\vspace{.5cm}
  \centering
  \includegraphics[width=.95\columnwidth]{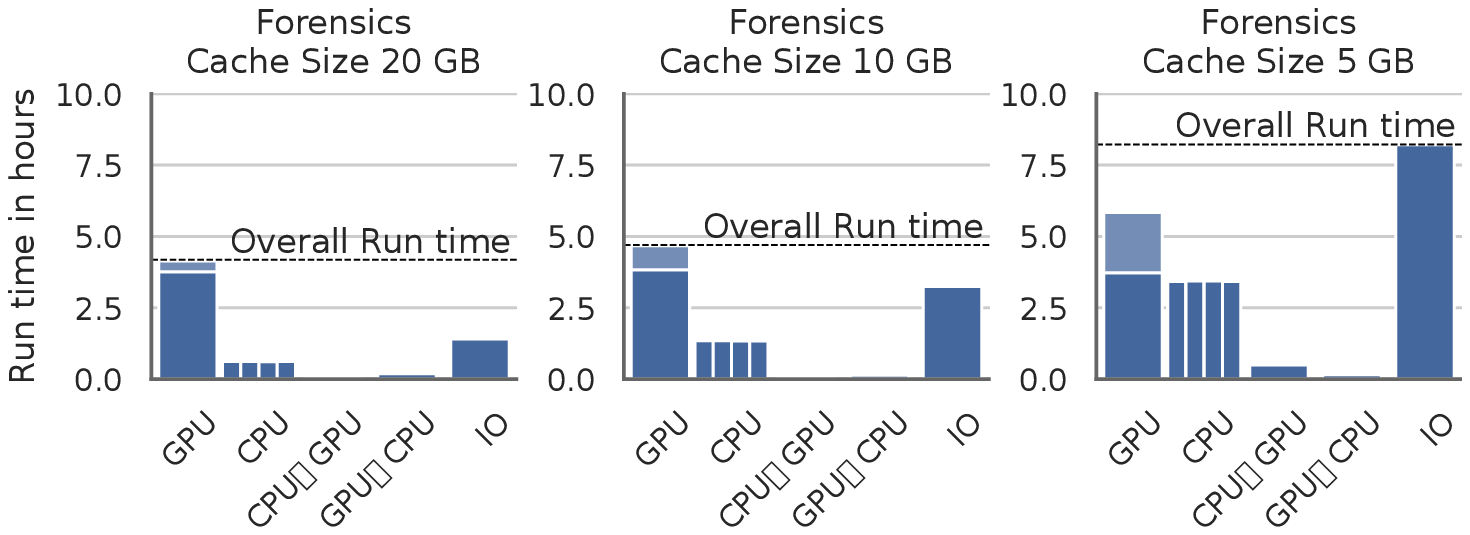}
  \caption{Processing time per thread on one node (NVIDIA TitanX Maxwell) for different host cache sizes. Results shown are for the forensics application.}
  \label{fig:experiment_cache_size_threads}
\end{figure}

To simulate a desktop computer with fewer available resources, we artificially further limit the number of local cache slots and study the effect on performance.
\Cref{fig:experiment_cache_size} shows performance when varying the number of cache slots (thus the maximum cache size $S$) of host and device cache.
For $S <11~\text{GB}$, the host cache was disabled and the device cache was set to $S$.
For $S > 11~\text{GB}$, the host cache was set to $S$ and the device cache to 11~GB (GPU memory capacity).

The microscopy application is not affected by the local cache size since its input easily fits into memory.
For the other two applications, the number of loads is inversely proportional to the local cache size and system efficiency gradually degrades when shrinking the cache.
For example, for the bioinformatics application, at 6~GB only 1.7\% of the inputs can be cached at any moment in time, but system efficiency is still 52.5\% compared to a hypothetical system having infinite memory.
Overall, Rockets continues to deliver decent performance even when limited to a tiny memory footprint.
This is due to the hierarchical processing approach which provides excellent data locality, even for a single node.

\Cref{fig:experiment_cache_size_threads} shows the processing time per thread when varying the local cache size for the forensics application.
The figure shows that decreasing the cache size results in increasing values $T_\text{CPU}$, $T_\text{GPU}$, and $T_\text{IO}$ since items are re-loaded more frequently.
On the other hand, increasing total cache capacity will thus lead to better performance and would be be possible by using more that one node.


\subsection{Scalability}
\label{sec:scalability}
\begin{figure}[t]
  \centering
  \includegraphics[width=.95\columnwidth]{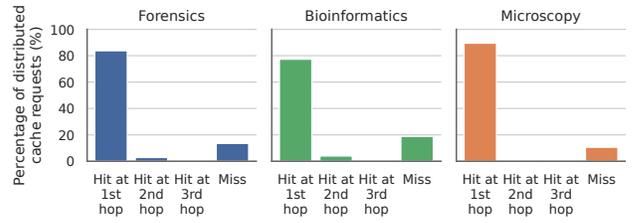}
  \caption{Percentage of cache hits/misses of the distributed cache $(h=3)$ for each of the three applications on 16 nodes (NVIDIA TitanX Maxwell).}
  \label{fig:experiment_cache_hops}
\end{figure}

In this section, we evaluate performance on 16 nodes, each having one NVIDIA TitanX Maxwell GPU.
First, we investigate parameter $h$ which specifies the maximum number of hops to check for each distributed cache request (see \cref{sec:design_cache_distributed}).
\Cref{fig:experiment_cache_hops}  shows the percentage of cache hits and misses for $h=3$.
The figure indicates that the vast majority of requests either results in a hit at the first hop (between $75{-}88\%$) or a miss (between $11{-}19\%$).
Subsequent hops after the first one thus contribute little to the number of cache hits.
The remaining experiments in this paper are performed for $h=1$ since this already provides an excellent cache hit ratio while generating the least amount of network traffic.

\begin{figure}[t!]
  \centering
  \includegraphics[width=.95\columnwidth]{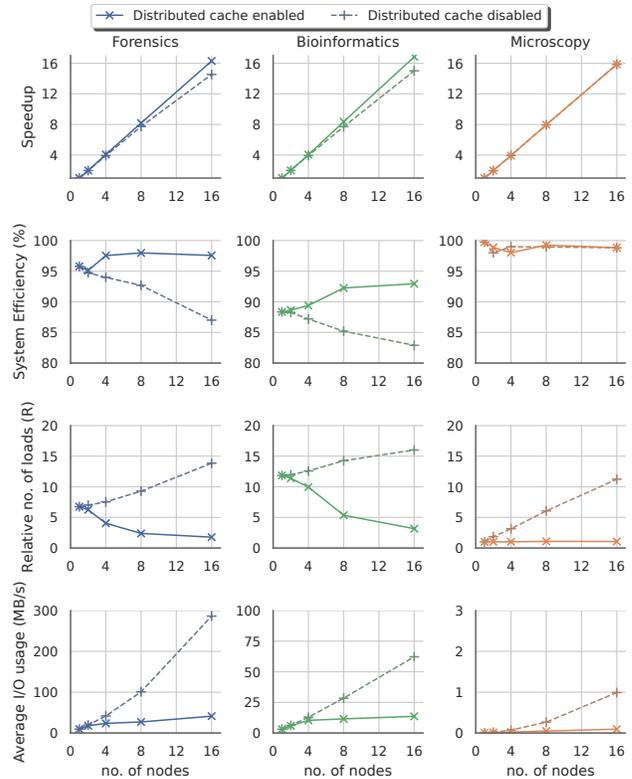}
  \caption{Speedup, system efficiency, factor $R$, and average I/O usage when scaling cluster size from 1 to 16 nodes (NVIDIA TitanX Maxwell).}
  \label{fig:experiment_scalability}
\end{figure}

We now consider scaling to 16 nodes for two scenarios: one with the distributed cache and one without.
\Cref{fig:experiment_scalability} shows speedup, system efficiency, data reuse, and average I/O usage versus the number of nodes for these two scenarios.
For the microscopy application, we see an excellent speedup of $15.8{\times}$ on 16 nodes; this application is expected to scale well since it is compute-intensive.
For the other two applications, we even see \emph{super}-linear speedup on 16 nodes when enabling the distributed cache ($16.9{\times}$ for bioinformatics and  $16.1{\times}$ for forensics), but \emph{sub}-linear speedup when disabling the distributed cache ($14.6{\times}$ for bioinformatics and $14.7{\times}$ for forensics).
The applications show much better scalability when the distributed cache is enabled.

The super-linear speedup can be understood when considering the total number of loads. 
The distributed cache exploits the larger combined memory capacity which results in better data reuse and thus higher efficiency.
For instance, for the forensics application, going from 1 to 16 nodes with the distributed cache means factor $R$ lowers from $6.7$ to $1.7$ and system efficiency increases from 95.8\% to 97.6\%.
Without, $R$ raises to $14.3$ and system efficiency decreases to $87.5\%$.

Besides run time, it is also important to consider the impact on I/O when scaling to large platforms: shorter run times combined with more nodes result in increased pressure on the storage system.
Most production platforms run more than one application simultaneously. 
Therefore, it is important to reduce the I/O pressure on the storage system, even if it does not directly improve the performance of our application. Less I/O pressure will improve the overall system performance.

\Cref{fig:experiment_scalability} shows that the average I/O usage (i.e., total bytes transferred by all nodes divided by total run time) is negligible for the microscopy application.
For the other two applications, I/O usage scales linearly with the number of nodes, but at a much slower rate when the distributed cache is enabled.
For instance, for the forensics application, the average I/O usage is 9.6~MB/s when using one node (137~GB over ${\sim}4$ hours).
Using 16 nodes with the distributed cache leads in I/O usage of only 39.9~MB/s, an increase of just $4.1{\times}$.
Disabling the distributed cache results in I/O usage of 294.7~MB/s (289~GB over 16.3 minutes), an increase of almost $31{\times}$ over one node.


\subsection{Heterogeneity}
\label{sec:evaluation_hetero}
Dynamic load-balancing means that Rocket can exploit heterogeneous systems efficiently, even if the applications are irregular or the system is shared with multiple users.
Nodes might also contain different GPUs from different generations, which is common in production environments since these platforms often replace GPUs in several phases throughout the lifetime of the system due to the fast evolution of GPUs.

\begin{figure}[t]
  \centering
  \includegraphics[width=.95\columnwidth]{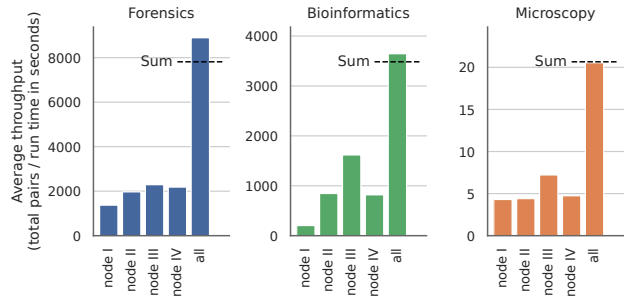}
  \caption{Results for heterogeneous runs of applications, see \cref{sec:evaluation_hetero}.}
  \label{fig:experiment_hetero}
\end{figure}

To demonstrate how our framework handles such a scenario, we execute the applications on four nodes equipped with different (combinations of) NVIDIA GPUs from different generations: 
node I (Kepler K20m), 
node II (Maxwell GTX980 + Pascal TitanX),
node III ($2 {\times}$Turing RTX2080Ti),
and node IV (Kepler Titan + Pascal TitanX).
\Cref{fig:experiment_hetero} shows the performance for each node individually and when using all four nodes together.
Performance is measured in average throughput (i.e, total pairs divided by total run time) to ease comparison of the performance.

\begin{figure}[t]
  \centering
  \includegraphics[width=.95\columnwidth]{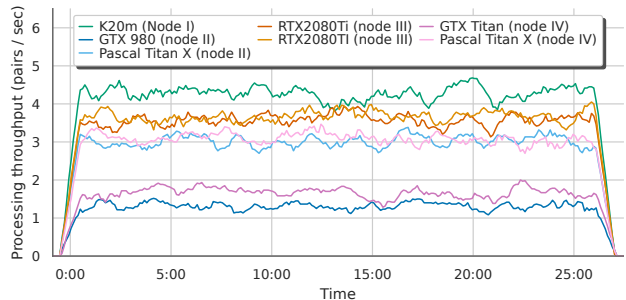}
  \caption{Heterogeneous run for the microscopy application, see \cref{sec:evaluation_hetero}. Throughput is measured using a rolling average of one minute. Fluctuations are due to the irregular run times of the pair computations, see \cref{fig:experiment_times}.}
  \label{fig:experiment_hetero_timeline}
\end{figure}

The results show good performance for each application on each of the four nodes individually, where some nodes inherently provide better performance (e.g., node III) than other nodes (e.g., node I) due to the performance differences of the GPUs.
Combining the four nodes should ideally provide performance equal to the sum of the individual nodes and the figure shows that the actual performance often even outperforms this sum due to the distributed cache.
Overall, Rocket delivers high performance even when on a diverse platform consisting of 7 GPUs from 4 different generations across 4 nodes.

\Cref{fig:experiment_hetero_timeline} shows the processing throughput over time (i.e., pairs processed per second) of the combined run for the microscopy application.
We make the following observations:
First, all nodes finish at roughly the same time, indicating that the workload is well-balanced.
Second, the processing rate is fairly consistent  across the run for each GPU, although fluctuations are present due to the irregularity of this application (i.e., some pairs take longer to process than others, see \cref{fig:experiment_times}).
Rocket is designed to always acquire more jobs well before the GPUs become idle, meaning the GPUs are always fully utilized.
Third,  the processing rate differs for the different devices, with more powerful GPUs (e.g., RTX2080Ti) delivering a higher processing rate than others (e.g., GTX980).


\begin{figure}[t]
  \centering
  \includegraphics[width=.9\columnwidth]{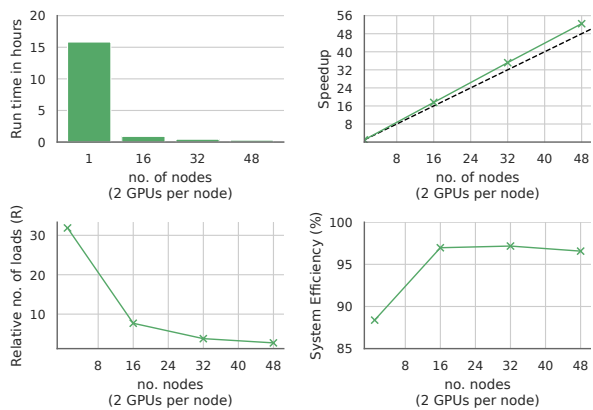}
  \caption{Results for the large-scale experiment on 96 GPUs (see \cref{sec:evaluation_hero}).}
  \label{fig:experiment_hero}
\end{figure}

\subsection{Large-scale Experiment}
\label{sec:evaluation_hero}

Large-scale experiments were performed on \emph{Cartesius} for the bioinformatics application since it requires the largest cache slot size, thus making it the most difficult application to maximize data reuse.
The input data set consists of all available reference bacteria proteomes (6818 files) from the Uniprot Proteomes database~\cite{2019uniprot} as of March 2020.

\Cref{fig:experiment_hero} shows the run time, speedup, $R$, and system efficiency when scaling from 1 node (2 GPUs) to 48 nodes (96 GPUs).
Run times decrease from 16 hours on one node to just under 20 minutes with 96 GPUs.
Super-linear speedup is present even on 96 GPUs which, as explained previously, is due to the distributed cache that exploits the larger combined memory capacity of the nodes.
The number of loads decreases by a factor $11.8{\times}$ from $R=31.9$ for 1 node to just $R=2.7$ for 48 nodes.




\section{Future Work}
\label{sec:future}

In this section we discuss several directions for future work.
We are working on extending the generality of our solution and cover more types of parallel applications that involve data reuse.
For example, applications with more complex workloads, such as processing triples (or any $n$-tuple) or using user-defined heuristics to reduce the number of pairs.
We are also considering applications that have more complex pipelines consisting of many phases and using different accelerators (e.g., FPGAs, APUs, co-processors).
An exciting direction is to extend the work-stealing algorithm with some form of cache-awareness such that remote tasks are chosen based on locally available data, thus enabling more reuse.

Furthermore, extending the caching design would also present many opportunities.
For example the ability to cache different items at different levels; persistent caches that reuse data from previous runs for the next execution; or including novel memory technologies (e.g., NVM, flash storage).
We are also working on solutions that enable variable-sized cache slots instead of fixed-sized ones.

Finally, other interesting system aspects we did not consider in this paper are fault-tolerance, energy consumption, elasticity, cloud environments, or multi-cluster computing.



\section{Conclusions}
\label{sec:conclusion}

In this paper, we have studied the problem of all-pairs compute problems.
Our solution combines multi-level caches to exploit data reuse; random work-stealing to allow dynamic workload balancing; a divide-and-conquer approach to exploit data locality; and asynchronous processing to overlap computation and data movement at all levels.
The implementation of our framework, \emph{Rocket}~\cite{rocket2020}, is available online.

We performed a detailed evaluation with three different real-world scientific applications on different platforms: from a single node, to a medium-scale heterogeneous cluster, and finally to a large-scale supercomputer containing 96 GPUs.
Results show that we achieve excellent scalability to multiple nodes, often showing super-linear speedup thanks to the distributed cache.
Moreover, we demonstrate perfect load balancing even on a highly heterogeneous platform.
Our results demonstrate, for example, that with Rocket we can reconstruct the evolutionary tree of all reference bacteria proteomes on Uniprot in under 20~minutes using a supercomputer.
We conclude that our Rocket framework is easy to use, and enables extremely efficient execution of all-pairs applications on large-scale systems, often achieving super-linear speedups.

\section*{Acknowledgment}
This project has received funding from 
the Netherlands eScience Center under file number 027.016.G06 (\emph{A methodology and ecosystem for many-core programming})
and
the \mbox{European} Union’s Horizon 2020 research and innovation programme under Grant Agreement 777533 (\emph{PROCESS}).


\IEEEtriggeratref{18}
\bibliographystyle{IEEEtran}
\bibliography{library}
\end{document}